\documentclass[reprint,aps,prl,showpacs,nofootinbib]{revtex4-1}
\usepackage{graphicx}
\usepackage{array}
\usepackage{amsmath}
\usepackage{epstopdf}
\newcommand{\EqLabel}[1]{\label{#1}}
\def\ie{{i.e}. } 
\begin{document}
\title{Efficient learning strategy of Chinese characters based on network approach}
\author{Xiaoyong Yan$^{1,2}$, Ying Fan$^{1,3}$, Zengru Di$^{1,3}$, Shlomo Havlin$^{4}$, Jinshan Wu$^{1,3,\dag}$}
\affiliation{1. Department of Systems Science, School of Management, Beijing Normal University, Beijing, 100875, P.R. China \\
2. Center for Complex Systems Research, Shijiazhuang Tiedao University, Shijiazhuang 050043, P.R. China \\
3. Center for Complexity Research, Beijing Normal University, Beijing, 100875, P.R. China \\
4. Department of Physics, Bar-Ilan University, Ramat-Gan 52900, Israel}
\begin{abstract}
Based on network analysis of hierarchical structural relations among Chinese characters, we develop an efficient learning strategy of Chinese characters. We regard a more efficient learning method if one learns  the same number of useful Chinese characters in less effort or time. We construct a node-weighted network of Chinese characters, where character usage frequencies are used as node weights. Using this hierarchical node-weighted network, we propose a new learning method, the distributed node weight (DNW) strategy, which is based on a new measure of nodes’ importance that takes into account both the weight of the nodes and the hierarchical structure of the network. Chinese character learning strategies, particularly their learning order, are analyzed as dynamical processes over the network. We compare the efficiency of three theoretical learning  methods and two commonly used methods from mainstream Chinese textbooks, one for Chinese elementary school students and the other for students learning Chinese as a second language. We find that the DNW method significantly outperforms the others, implying that the efficiency of current learning methods of major textbooks can be greatly improved.
\end{abstract}
\date{\today}
\maketitle

{\bf{Introduction}}. It is widely accepted that learning Chinese is much more difficult than learning western languages, and the main obstacle is learning to read and write Chinese characters. However, some students who have learned certain amount of Chinese characters and gradually understand the intrinsic coherent structure of the relations between Chinese characters,  quite often find out that it is not that hard to learn Chinese \cite{Bellassen}. Unfortunately, such experiences are only at individual level. Until today there is no textbook that have exploited systematically the intrinsic coherent structures to form a better learning strategy. We explore here such relations between Chinese characters systematically and use this to form an efficient learning strategy.

Complex networks theory has been found useful in diverse fields, ranging from social systems, economics to genetics,  physiology and climate systems \cite{Watts, Strogatz, Albert, Newman, Wu, Costa, Fortunato}. An important challenge in studies of complex networks in different disciplines is how network analysis can improve our understanding of function and structure of complex systems \cite{Costa, Fortunato, Chen}. Here we address the question if and how network approach can improve the efficiency of Chinese learning.

Differing from western languages such as English, Chinese characters are non-alphabetic but are rather ideographic and orthographical \cite{Branner}. A straightforward example is the relation among the Chinese characters `\raisebox{-1mm}{\includegraphics[width=0.5cm]{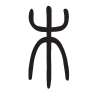}}', `\raisebox{-1mm}{\includegraphics[width=0.5cm]{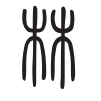}}' and `\raisebox{-1mm}{\includegraphics[width=0.5cm]{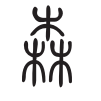}}', representing  tree, woods and forest, respectively. These characters appear as one tree, two trees and three trees. The connection between the composition forms of these characters and their meanings is obvious. Another example is `\raisebox{-1mm}{\includegraphics[width=0.5cm]{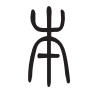}}' (root), which is also related to the character ` \raisebox{-1mm}{\includegraphics[width=0.5cm]{char-tree.pdf}}' (tree):  A bar near the bottom of a tree refers to the tree root. Such relations among Chinese characters are common, though sometimes it is not easy to realize them intuitively, or, even worse, they sometimes may become fuzzy after a few thousand years of evolution of  the Chinese characters. However, the overall forms and meanings of Chinese characters are still closely related \cite{Qiu, Bai, Bellassen}: Usually, combinations of simple Chinese characters are used to form complex characters. Most Chinese users and learners eventually notice such structural relations although quite often implicitly and from accumulation of knowledge and intuitions on Chinese characters \cite{Lam1}. Making use of such relations explicitly might be helpful in turning rote leaning into meaningful learning \cite{Novak:Cmap}, which could improve efficiency of students' Chinese learning. In the above example of `\raisebox{-1mm}{\includegraphics[width=0.5cm]{char-tree.pdf}}', ` \raisebox{-1mm}{\includegraphics[width=0.5cm]{char-woods.pdf}}', and `\raisebox{-1mm}{\includegraphics[width=0.5cm]{char-forest.pdf}}', instead of memorizing all three characters individually in rote learning, one just needs to memorize one simple character `\raisebox{-1mm}{\includegraphics[width=0.5cm]{char-tree.pdf}}' and then uses the logical relation among the three characters to learn the other two.

However, such structural relations among Chinese characters have not yet been fully exploited in practical Chinese teaching and learning. As far as we know from all mainstream Chinese textbooks the textbook of Bellassen et al. \cite{Bellassen} is the only one that has taken partially the structure information into consideration. However, considerations of such relations in teaching Chinese in their textbook are, at best, at the individual characters level and focus on the details of using such relations to teach some characters one-by-one. With the network analysis tool at hand, we are able  to analyze this relation at a system level. The goal of the present manuscript is to perform such a system-level network analysis of Chinese characters and to show that it can be used to significantly improve Chinese learning.

Major aspects of strategies for teaching Chinese include character set choices, the teaching order of the chosen characters, and details of how to teach every individual character. Although our investigation is potentially applicable to all three aspects, we focus here only on the teaching order question. Learning order of English words is a well studied question which has been well established \cite{English_Order}. However, there is almost no explicit such studies in Chinese characters. In this work, the characters choice is taken to be the set of the most frequently used characters, with $99\%$ accumulated frequency \cite{Frequency}. To demonstrate our main point: how network analysis can improve Chinese learning, we focus here on the issue of Chinese character learning order.

Although some researchers have applied complex network theory to study the Chinese character network \cite{Li, Lee}, they mainly focus on the network's structural properties and/or evolution dynamics, but not on learning strategies.  A recent work studied the evolution of relative word usage frequencies and its implication on coevolution of language and culture \cite{Petersen}. Different from these studies, our work considers the whole structural Chinese character network, but more importantly, the value of the network for developing efficient Chinese characters learning strategies. We find, that our approach,  based on both word usage and network analysis provides a valuable tool for efficient language learning.

{\bf{Data and methods.}} Although nearly a hundred thousand Chinese characters have been used throughout history, modern Chinese no longer uses most of them. For a common Chinese person, knowing $3,000 - 4,000$ characters will enable him or her to read modern Chinese smoothly. In this work, we thus focus only on the most used $3500$ Chinese characters, extracted from a standard character list provided by the Ministry of Education of China  \cite{Characters}. According to statistics \cite{Frequency}, these 3500 characters account for  more than $99\%$ of the accumulated usage frequency in the modern Chinese written language.

\begin{figure}
\includegraphics[width=8.4cm]{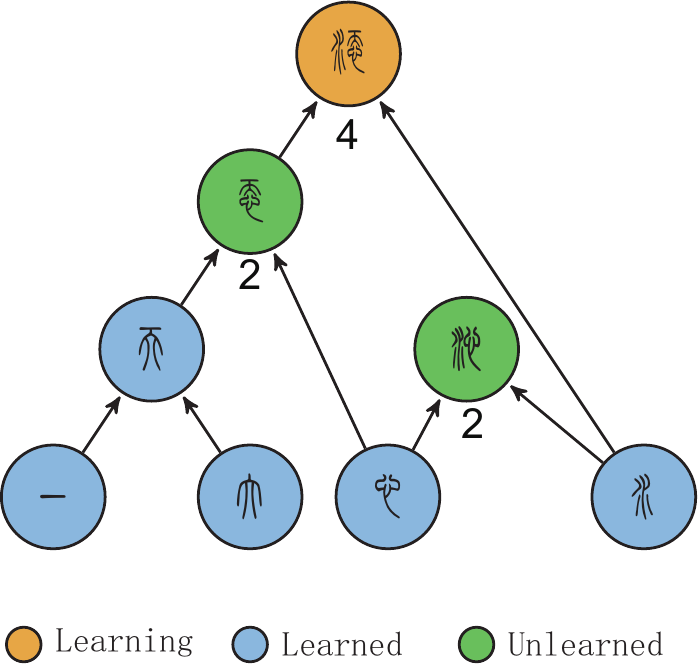}
\caption{\label{fig1} Chinese character decomposing and network construction. The numerical values in the figure represent learning cost, which will be discussed later.}
\end{figure}

Most Chinese characters can be decomposed into several simpler sub-characters \cite{Qiu,Bai}. For instance, as illustrated in Fig. \ref{fig1}, character `\raisebox{-1mm}{\includegraphics[width=0.5cm]{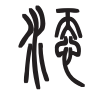}}'(means `add') is made from `\raisebox{-1mm}{\includegraphics[width=0.5cm]{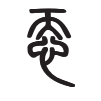}}'(ashamed) and `\raisebox{-1mm}{\includegraphics[width=0.5cm]{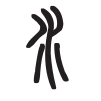}}'(water); `\raisebox{-1mm}{\includegraphics[width=0.5cm]{char-shame.pdf}}' can then be decomposed into `\raisebox{-1mm}{\includegraphics[width=0.5cm]{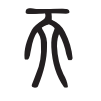}}'(head, or sky) and `\raisebox{-1mm}{\includegraphics[width=0.5cm]{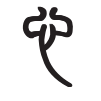}}'(heart), and `\raisebox{-1mm}{\includegraphics[width=0.5cm]{char-sky.pdf}}' can be decomposed into `\raisebox{-2mm}{\includegraphics[width=0.5cm]{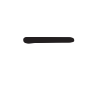}}' (one) and `\raisebox{-1mm}{\includegraphics[width=0.5cm]{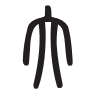}}'(a person standing up, or big). The characters `\raisebox{-1mm}{\includegraphics[width=0.5cm]{char-water.pdf}}', `\raisebox{-1mm}{\includegraphics[width=0.5cm]{char-heart.pdf}}', ` \raisebox{-1mm}{\includegraphics[width=0.5cm]{char-one.pdf}}' and `\raisebox{-1mm}{\includegraphics[width=0.5cm]{char-big.pdf}}' cannot be decomposed any further, as they are all radical hieroglyphic symbols in Chinese.  There are general principles about how simple characters form compound characters. It is so-called ``Liu Shu'' (six ways of creating Chinese characters). Ideally when for example two characters are combined to form another character the compound character should be connected to its sub-characters either via their meanings or pronunciations. We have illustrated those principles using characters listed in Fig. \ref{fig1}. See {\bf{Supporting Online Material}} for more details. While certain decompositions are structurally meaningful and intuitive,  others are not that obvious at least with the current Chinese character forms \cite{Bai}. In this work, we do not care about the question, to what extent Chinese character decompositions are reasonable, the so-called Chinese character rationale \cite{Qiu}, but rather about the existing structural relations (sometimes called character-formation rationale or configuration rationale) among Chinese characters and how to extract useful information from these relations to learn Chinese. Our decompositions  are based primarily on Ref. \cite{ShuoWen, Qiu, Bai}.

\begin{figure}
\includegraphics[width=8.4cm]{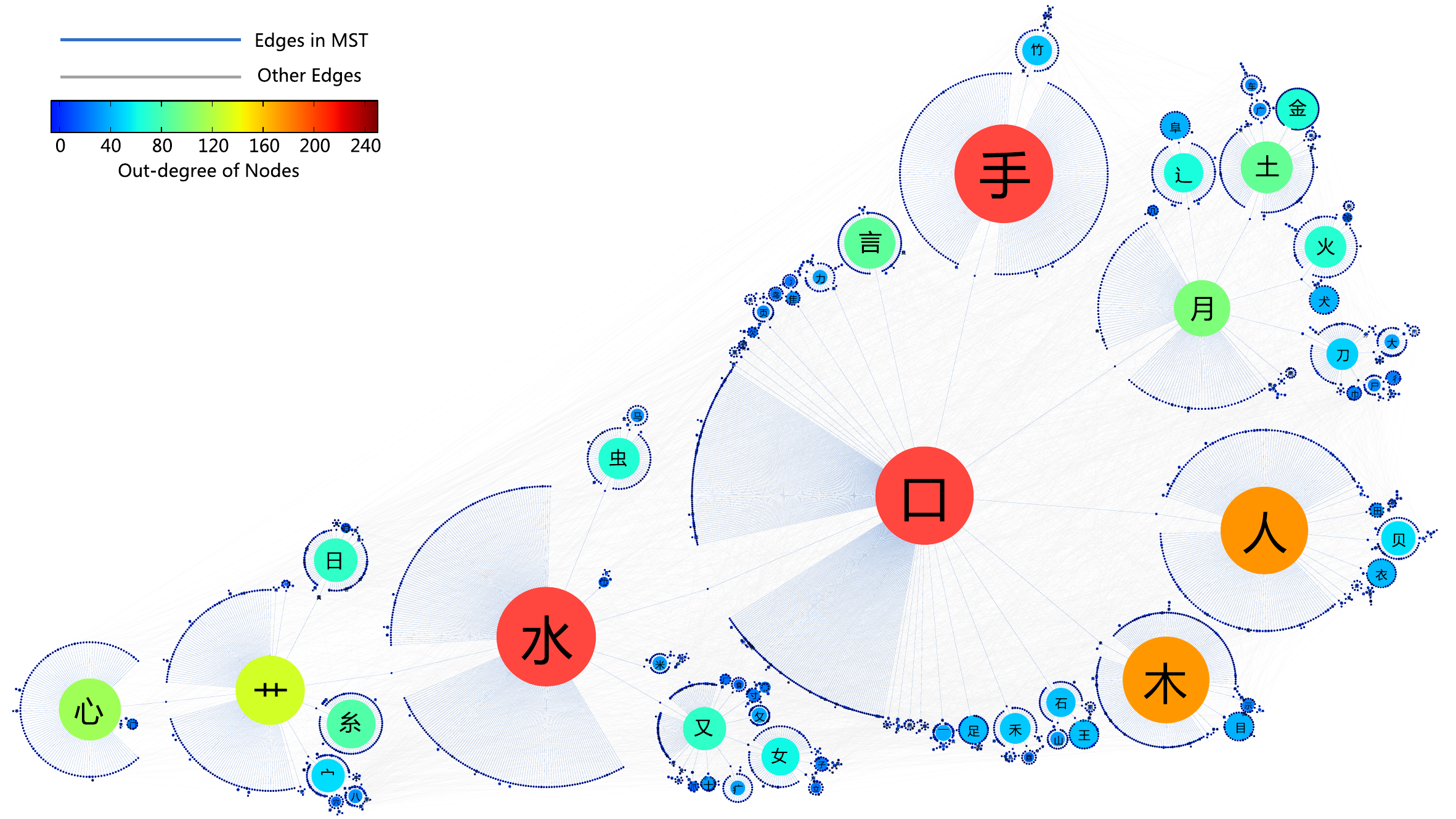}
\caption{\label{fullmap} Full map of the Chinese character network. For a better visual demonstration, we plot here the minimum spanning tree of the whole network which is shown in blue while other links are presented in  grey as a background. All characters can be seen when the figure magnified properly. }
\end{figure}

Following the general principles shown in the above example and the information in Ref. \cite{ShuoWen,Qiu, Bai} , we decompose all 3500 characters and construct a network by connecting character $B$ to $A$ (an adjacent matrix element $a_{BA}=1$, otherwise it is zero) through a directed link if $B$ is a ``direct'' component of $A$. Here, ``direct'' means to connect characters hierarchically  (see Fig. \ref{fig1}):  Assuming $B$ is part of $A$, if $C$ is part of $B$ and thus in principle $C$ is also part of $A$, we connect only $B$ to $A$ and $C$ to $B$, but NOT $C$ to $A$.  There are other considerations on including more specific characters which are not within the list of most-used $3500$ characters but are used as radicals of characters in the list, in constructing this network. More technical details can be  found in the {\bf{Supporting Online Material}}. Decomposing characters and building up links in this way, the network is a Directed Acyclic Graph (DAG), which has a giant component of  $3687$ nodes (see  {\bf{Supporting Online Material}} for details on the number of nodes) and $7024$ links, plus $15$ isolated nodes. Fig. \ref{fullmap} is a skeleton illustration of the full map of the network.

\begin{figure}
\includegraphics[width=8.4cm]{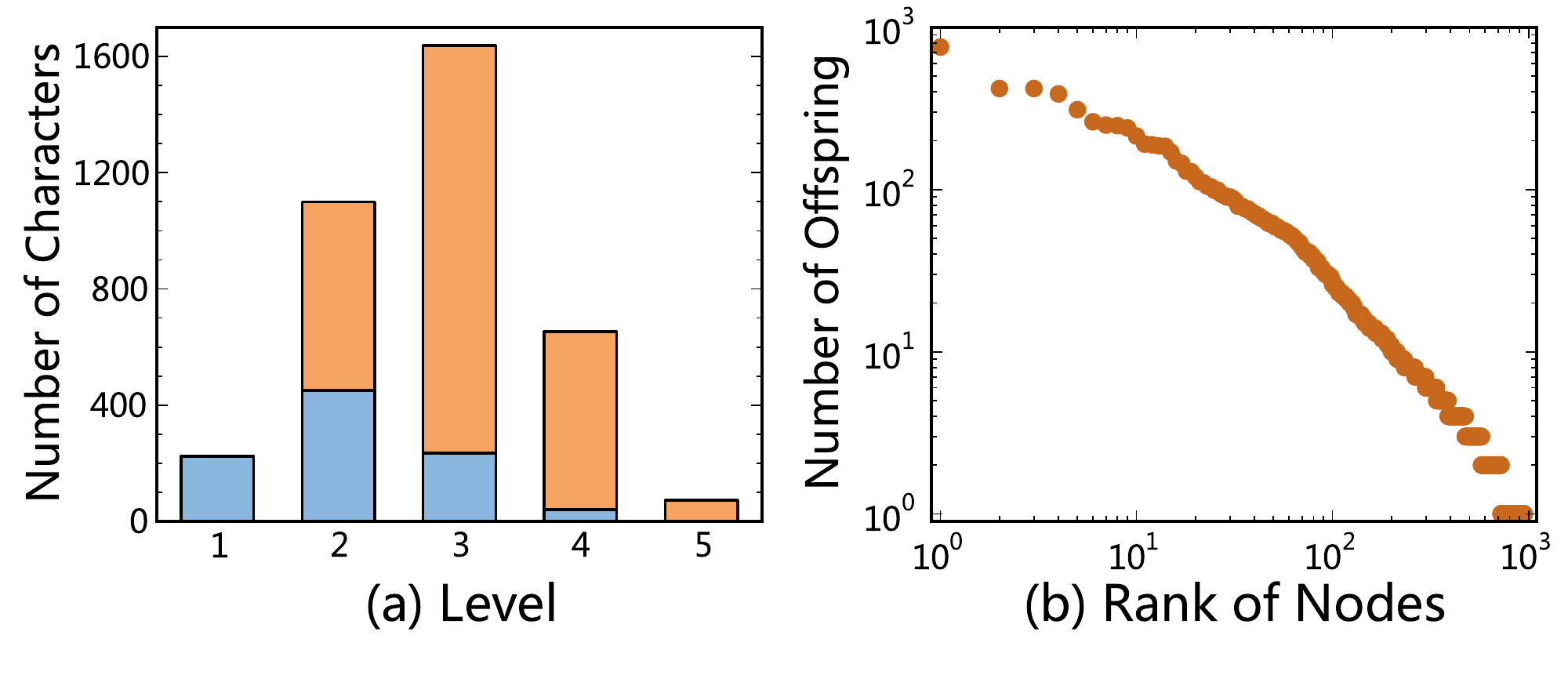}
\caption{\label{fig2} Topological properties of Chinese character network. (a) Hierarchical distribution: number of characters at each level. The number of characters in each level that have no offspings is shown in brown. (b) Node-offspring distribution: Zipf plot, where characters are ranked according to their number of offsprings. The number of offsprings of a character is plotted against the rank of the character.}
\end{figure}

As a DAG, the Chinese character network is hierarchical. Starting from the bottom in Fig. \ref{fig1}, where nodes have no incoming links, we can assign a number to a character to denote its level: all components of a character should have lower levels than the character itself. Fig.\ref{fig2}(a) shows the hierarchical distribution of characters in the network. The figure shows that the network has a small set of radical characters ($224$ nodes at the bottom level, $1$) and nearly $94\%$ of the characters lie at higher levels. Moreover, the network has a broad heterogeneous offsprings degree distribution (a node's offspring degree is defined as its number of outgoing edges). Notice in Fig. \ref{fig2}(b), the number of characters with more than one (the smallest number on the vertical axis) offspring is close to $1000$ (the largest number shown on the horizontal axis). This means that less than $1000$ of the $3687$ characters are involved in forming other characters. The other characters are simply the top ones in their paths so that no characters are formed based on them. Their distribution in the different levels is also shown in Fig. \ref{fig2}a.

{\bf{Learning Strategy.}} The heterogeneity of the hierarchical structure reflected in the node-offspring broad distribution in the Chinese character network suggests that learning Chinese characters in a ``bottom-up" order (starting from level $1$ characters and gradually climbing along the hierarchical paths) may be an efficient approach. At the level of learning of {\it{individual}} characters, Chinese teaching has indeed used this rationale\cite{Bellassen, Zhou}. Other approaches are based on character usage frequencies, \ie learning the most used characters, \ie those appearing as the most used words first (Ref.  \cite{Lam2} provides a critical review of this approach and others).

To assess the efficiency of different approaches, which is here limited to Chinese characters learning orders, one needs a method to measure the learning efficiency. However, measuring learning efficiency is not trivial and currently, to the best of our knowledge, does not exist. In our approach, we regard a learning strategy as more efficient if it reaches the same learning goal, \ie a desired number of learned characters or accumulated character usage frequencies, with lower learning costs compared to other strategies.

The question thus becomes how to determine the learning cost? Of all possible factors related to cost, it is reasonable to assume that a character with more sub-characters and more unlearned sub-characters is more difficult to learn. For example, the character `\raisebox{-1mm}{\includegraphics[width=0.5cm]{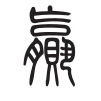}}', with 5 sub-characters, is obviously more difficult to learn than `\raisebox{-1mm}{\includegraphics[width=0.5cm]{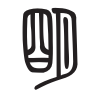}}', with 2 sub-characters. Conversely, it is easier to learn a character for which all sub-characters have been learned earlier than another character with same number of sub-characters all of which are previously unknown to the learner. We thus intuitively define the cost for a student to learn a character as the sum of the number of sub-characters and the learning cost of the unlearned sub-characters at his current stage.  The learning cost of the unlearned sub-characters is calculated recursively until characters at the first level are reached or until all sub-characters have been learned previously. Each unlearned character of the first level contributes cost $1$, while previously learned characters contribute cost $0$. For example, assuming that, at a given stage, a student needs to learn the character `\raisebox{-1mm}{\includegraphics[width=0.5cm]{char-add.pdf}}' and that the student already knows the characters in blue in Fig. \ref{fig1}. We demonstrate the cost for the student to learn this character. First, the character ` \raisebox{-1mm}{\includegraphics[width=0.5cm]{char-add.pdf}}' has $2$ sub-characters (`\raisebox{-1mm}{\includegraphics[width=0.5cm]{char-water.pdf}}'and `\raisebox{-1mm}{\includegraphics[width=0.5cm]{char-shame.pdf}}'), and the student does not know one character, `\raisebox{-1mm}{\includegraphics[width=0.5cm]{char-shame.pdf}}'. The total cost of learning the character `\raisebox{-1mm}{\includegraphics[width=0.5cm]{char-add.pdf}}' is thus equals to $2$ plus the cost of learning `\raisebox{-1mm}{\includegraphics[width=0.5cm]{char-shame.pdf}}', which, calculated using the same principle, is $2$ ($2$ sub-characters `\raisebox{-1mm}{\includegraphics[width=0.5cm]{char-sky.pdf}}' and `\raisebox{-1mm}{\includegraphics[width=0.5cm]{char-heart.pdf}}' , and none of which are new to the student). The cost for the student is thus $4$. If the student somehow learned the character `\raisebox{-1mm}{\includegraphics[width=0.5cm]{char-shame.pdf}}' before and then needs to learn `\raisebox{-1mm}{\includegraphics[width=0.5cm]{char-add.pdf}}', the cost of acquiring `\raisebox{-1mm}{\includegraphics[width=0.5cm]{char-add.pdf}}' is only $2$. Thus, to learn both characters, it is cheaper to first learn ` \raisebox{-1mm}{\includegraphics[width=0.5cm]{char-shame.pdf}}' and then `\raisebox{-1mm}{\includegraphics[width=0.5cm]{char-add.pdf}}' (total cost $2+2=4$), rather than the other way around ($4+2=6$).

\begin{figure}
\includegraphics[width=4.2cm]{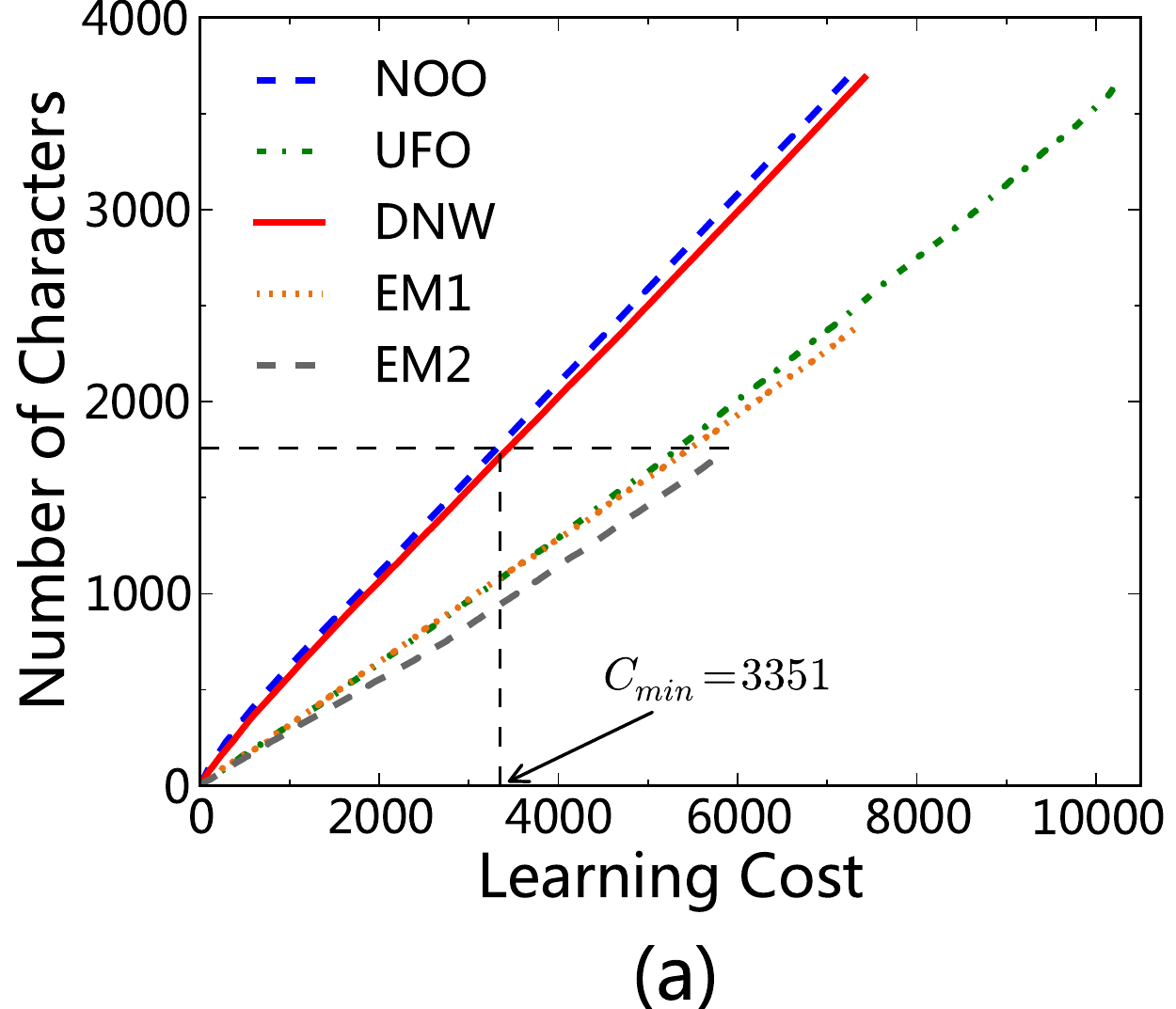}
\includegraphics[width=4.2cm]{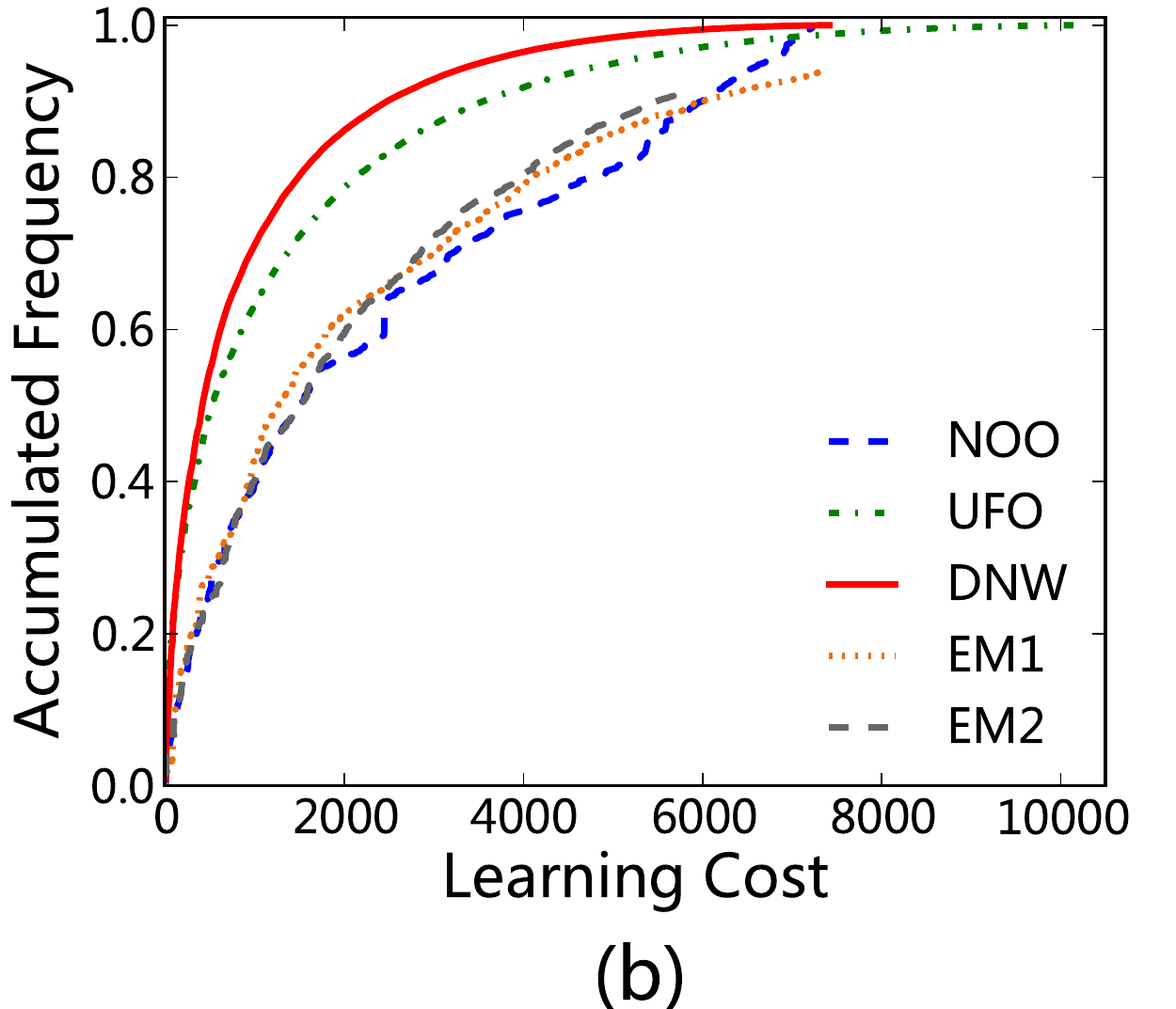}
\caption{\label{fig3} Learning efficiency comparison for different learning orders: node-offspring order (NOO), usage frequency order (UFO), distributed node weight (DNW) and two common empirical orders (EM1 for Chinese pupils and EM2 for LCSL). (a) Number of characters is set as the learning goal. (b) Accumulated usage frequency is set as the learning goal. $C_{min}$ is defined as the learning cost of $1775$ characters using the NOO method and it will be used in discussion of leaning efficiency index.}
\end{figure}

If we assume that learning more characters, independent of their usage frequency,  is the learning goal, the optimal learning strategy is to follow the node-offspring order (NOO) from many to few, which means learning characters with more offspring first. In this way,  an ancestor character is always learned before its offspring characters since the ancestor has at least one more offspring than the offspring character. From the learning cost definition, we know that using this approach we never waste effort in learning characters twice. No other strategy is thus better than this one. However,  in this way we might learn many characters with low usage frequencies which are less useful. Hence, as shown in Fig. \ref{fig3}b, if our aim is acquiring more accumulated usage frequency, the NOO-based strategy is indeed not a good one. Being able to achieve a high accumulated usage frequency in relatively short times is not only good for those who can not spend much time but it will also help the students to do extracurricular reading.

Thus, our main objective is to develop a learning strategy that reaches the highest accumulated usage frequency with limited cost. When simply following the character usage frequency order (UFO method) from high to low, one discards topological relations among characters that could help in the learning process and save cost. In UFO one learns characters at higher levels before learning those at lower levels, which is more costly. Thus, the question comes to developing a new Chinese character centrality measure of character importance, that considers both topological relations and usage frequencies. Such a measure could help to obtain a learning order better than both NOO and UFO. One additional consideration is to learn first the characters with larger out degree in the character network since here a large out degree means the character is involved as a component in many characters. The method proposed in the following in fact takes all these three aspects into consideration.

Here we develop a centrality measure that we call distributed node weight (DNW) based on both network structure and on usage frequencies which are the node weights ($W^{(m)}_j$ ). Here $j$ represents the node (character) and $m$ its level in the network.  The top level is $m=5$ (no outgoing links) and the bottom level is $m=0$ (no incoming links). To measure character centrality of node $j$ at level $m$, we pick each of its predecessors (denoted as node $i$ at level $m+1$) and add its weight $W^{(m+1)}_i$ multiplied by $b$ to the weight $W^{(m)}_j$ as follow:
\begin{equation}
\label{eq1}
\tilde{W}^{(m)}_j=W^{(m)}_j+b\sum_{i}W^{(m+1)}_i a_{ji},
\end{equation}
where $b\geq0$ is a parameter, $a_{ji}=1 \mbox{ or } 0$ is the adjacency matrix element from node $j$ to node $i$ (\ie whether or not character $j$ is a direct part of character $i$). In  the DNW method one learns characters in order according to their centrality  from highest to lowest.  Thus,  when $b=0$, the DNW is equivalent to the UFO method. For $b>0$, the node's offsprings play an important role. When $b=1$ and all $W_{j}=1$ (which means ignoring the difference in character usage frequencies), the DNW centrality order becomes the node-offspring order (NOO). In this sense, the NOO is an unweighted version of the DNW. The DNW order can thus be considered a hybrid of the NOO and UFO.

Using numerical analysis, we find that the optimal $b$ value for the DNW strategy is $b\simeq 0.35$, as discussed below. With this optimal parameter $b$, we compare our strategy of DNW learning order against the NOO and the UFO in Fig.\ref{fig3}. We find in Fig.\ref{fig3}a that DNW is close to NOO, regarding the total number of characters vs. the learning cost. However, in Fig. \ref{fig3}b,  the DNW is significantly better than NOO and even better than UFO, regarding the total accumulated usage frequency vs. the learning cost. In the left panel,  NOO and DWN are much better than UFO, while in the right panel the UFO and DNW are much better than NOO. Thus, only the DNW demonstrates a high efficiency in both, accumulated frequency and total number of characters.


The DNW in the right figure appears to be only slightly better than the UFO, but this is a little misleading. From the left figure, we can see that with the same cost, say around $1000$, although the difference between the two is relatively small in the right figure, there is a much bigger difference in the left figure. It means that even though the DNW is only slightly better than the UFO on the accumulated usage frequency, significantly more characters are learned following the DNW than the UFO. Such a difference in number of known characters sometimes is as important as the accumulated usage frequency when estimating if an individual is literate or not. For beginners, $400-500$ characters is roughly the first barrier. Many stop there. Using the UFO, this corresponds to a cost of about $2000$ while using the DNW it is around only $1000$. Thus, it will be much easier for students to overcome this barrier when using DNW compared to UFO.

We next compare the DNW against two empirical commonly used orders: one is from a set of the most used Chinese textbook \cite{Textbook1} for primary schools in China, which contains $2475$ different Chinese characters (EM1); the other is from a mainstream Chinese textbook \cite{Textbook2} for students Learning Chinese as a Second Language (LCSL), which contains $1775$ different Chinese characters (EM2). We sort the two character sets by  first appearances in new character lists in the two textbooks and plot their learning results in Fig.\ref{fig3}. The figure shows that compared to our developed DNW method, the empirical learning orders have relatively poor performance in both the total number of characters and accumulated usage frequency. This emphasizes the urgent need of improving the efficiency of current learning Chinese characters.

{\bf{Optimal b.}}  To find the optimal $b$ value, we define an efficiency index for learning strategies. We first take a certain learning cost and denote it as $C_{min}$, which is here set to be the learning cost of learning the total of $N_{min}=1775$ characters using the NOO order ($C_{min}=3351$, See Fig. \ref{fig3}a). We intuitively assume that the sooner a curve reaches $N_{min}$ the learning is more efficient. Thus, the larger is the area under the curves in Fig. \ref{fig3}a the learning can be regarded as more efficient. The same consideration holds for the curves in Fig. \ref{fig3}b. We therefore, measure the area underneath the learning efficiency curves (Fig.\ref{fig3}) up to cost $C_{min}$ and denote them as $S_n$ (area under the curve of number of characters v.s. cost like the ones in Fig. \ref{fig3}a) and similarly $S_f$ (area under the curve of accumulated usage frequency v.s. cost like those in Fig. \ref{fig3}b), respectively. The ratio between the area underneath the curves $S_{n}$ ($S_{f}$) and the area of a rectangular region defined by $C_{min}N_{min}$ ($C_{min}F_{min}$, where $F_{min}$ is the maximum accumulated frequency of the curves at $C=C_{min}$) is defined as the learning efficiency index,
\begin{align}
v_n=\frac{S_n}{C_{min}N_{min}},\\
v_f=\frac{S_f}{C_{min}F_{min}}.
\EqLabel{eq:speed}
\end{align}
The sooner a curve reaches $N_{min}$ ($F_{min}$) the larger is the area and so is the ratio, the more efficient is the learning order. In this sense, the above ratios serve as indexes of efficiency of learning orders.

\begin{figure}
\includegraphics[width=8.4cm]{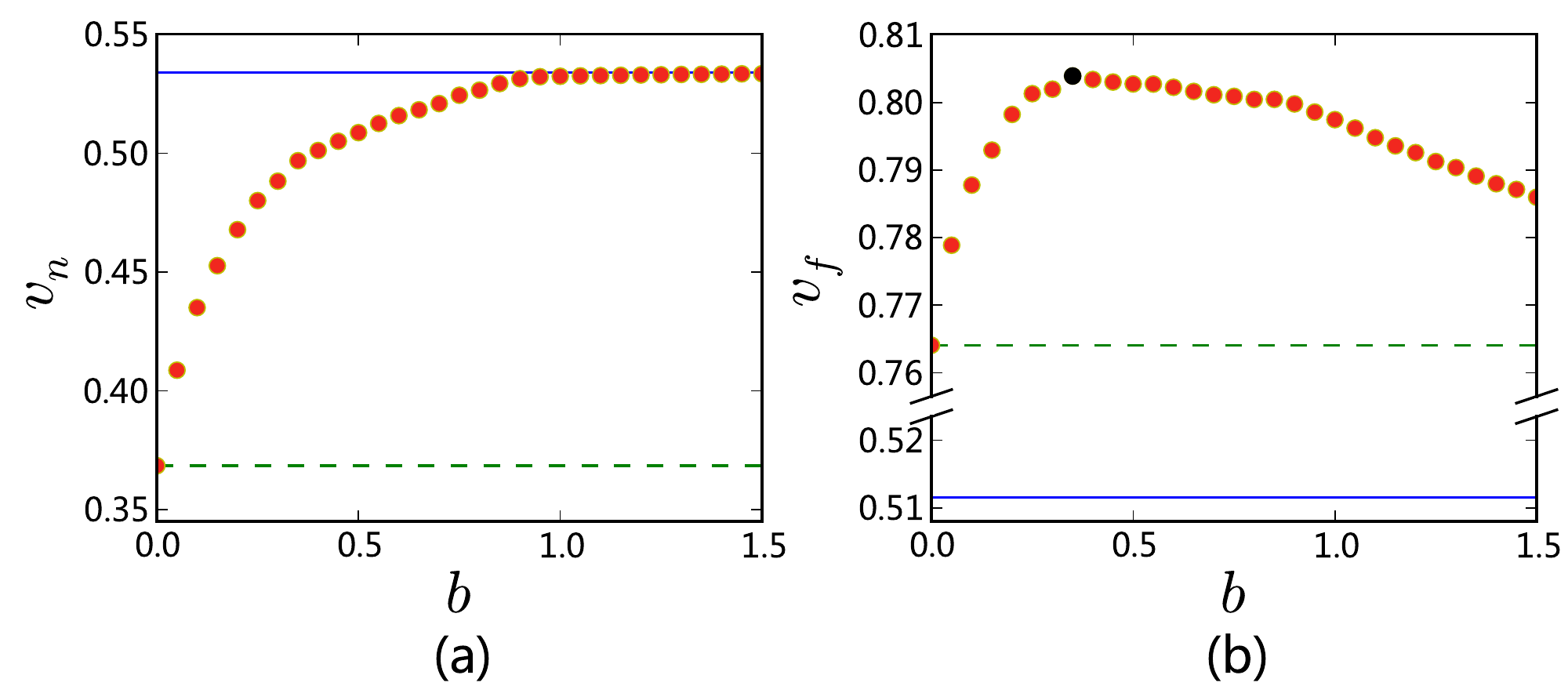}
\caption{\label{fig4} Efficient index of hybrid strategies as a function of b (dots). The two horizontal lines are the efficiency of the node-offspring order (blue line) and usage frequency order (green line). (a)  Efficiency when using number of characters as the learning goal. (b) Efficiency  when using accumulated usage frequency as the learning goal.}
\end{figure}

In Fig. \ref{fig4}, we plot $v_n$ and $v_f$ of the hybrid strategy (DNW) as functions of $b$. We also plot two lines, for comparison, showing the learning efficiency of the NOO (blue line) and UFO (green line). As $b$ increases, $v_n$ of the hybrid strategy approaches that of the NOO. On the other hand, when $b=0.35$, $v_f$ of hybrid strategy reaches its maximum. Thus, with respect to frequency usage the DNW with $b=0.35$ is the most efficient. However, if we consider also the number of characters the range of $b\in\left[0.35, 0.7\right]$ can be regarded as very good choices. As an example, in this work we use $b=0.35$, which shows a significant improvement over commonly used methods (Fig. \ref{fig3}).

\begin{figure}
\includegraphics[width=8.4cm]{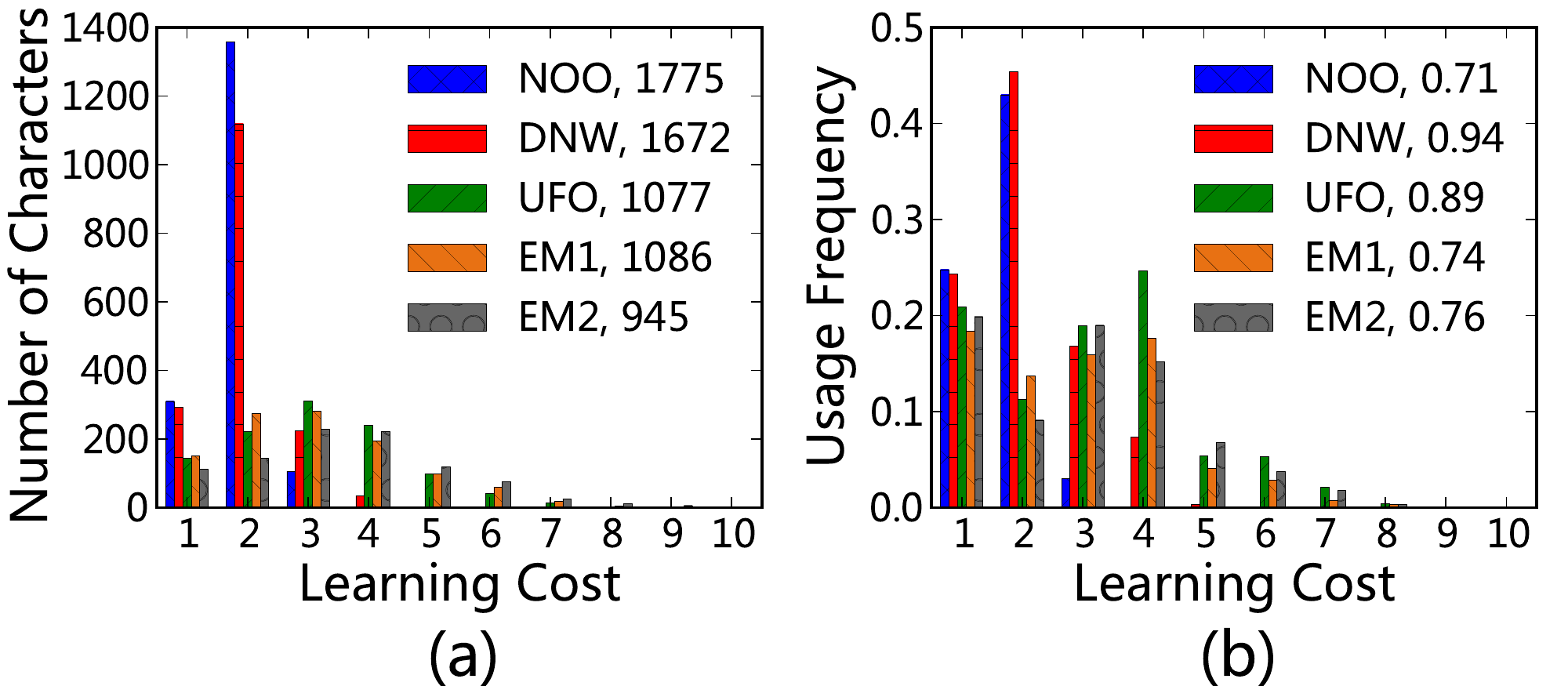}
\caption{\label{fig5} Up to a fixed total learning cost $C_{min}$, for all five learning orders, we count and plot the number of characters according to their individual learning costs in (a) and convert the number of characters into the corresponding usage frequency in (b). }
\end{figure}

In order to compare the DNW strategy against others in more detail, we have analyzed the learning cost statistics of the characters covered by cost $C_{min}$ for all the five learning strategies in Fig. \ref{fig5}. Recall that $C_{min}$ is the cost of learning first $1775$ characters using the NOO and number of characters covered by this $C_{min}$ is different for different methods.  Using the measure of learning cost proposed earlier, we record the learning cost of every character before the accumulated cost reaches $C_{min}$ in each learning order and then plot a histogram of learning costs of all those characters for each learning order. From Fig. \ref{fig5}a, we see that in both DNW and NOO learning orders, characters with learning cost $2$ are dominant (roughly $80\%$). In these two learning orders, few characters have learning cost higher than $3$. The other three learning orders have much smaller fraction of characters of cost-$2$ and more characters with cost higher than $3$. Most Chinese characters can be decomposed into $2$ direct parts, therefore, learning cost $2$ means that when a character is learned, its parts have been quite often learned before. This is natural in the NOO order since it is designed that way. However, as seen here it also holds in the DNW order, which is the high advantage of the DNW order. In Fig. \ref{fig5}b we also plot the corresponding usage frequencies of the set of characters with the same learning cost. In DNW one learns in fact about 6$\%$ less characters compared to NOO, but the usage of the characters learned in DNW is more than 30$\%$ higher. Thus DNW is significantly better than NOO. We also find that although DNW and UFO have comparable overall usage frequencies, the DNW is concentrated on the cost-$1$ and cost-$2$ characters while the UFO is distributed widely on characters with learning cost from $1$ to $4$. This illustrates further why our DNW is an efficient learning order in both the sense of total number and total usage frequency of characters.

{\bf{Conclusion and Discussion}}. We demonstrate the potential of network approach in increasing significantly the efficiency of learning Chinese. By including character usage frequencies as node weights to the structural character network, we discover and develop an efficient learning strategy which enables to turn rote learning of Chinese characters to meaningful learning. In the {\bf{Supporting Online Material}}, we present an adjacency list form of the constructed network; we also list Chinese characters order according to our DNW centrality. The constructed network might also help design a customized Chinese character learning order for students who have previously learned some Chinese and want to continue their studies at their own paces. Given the information about the student's known characters in our network, our DNW centrality measure can be adapted to be used in finding a specific student oriented optimal learning order. This goal is completely out of reach of standard textbook-based education and it will be especially useful for Chinese learners that do not study Chinese in a formal Chinese school, or study Chinese every now and then or using private tutors. We hope that our study will lead to develop textbooks applying the DNW learning order and detailed decomposition of each character. It will also be valuable for Chinese learners to have a dictionary explaining every character and word simply from a core set of small number of basic characters. Note that we are not claiming that our decomposition is perfect or that our character choice is good enough. These questions are still debated in the Chinese character structure fields. There are possibly also other topological quantities that might be valuable for Chinese learning. Considering our node-weighted network, the concept of using the shortest path to accumulate the largest node weight in shortest steps, clearly differs from the usual shortest path. How these quantities are related to Chinese learning is an interesting question that we have not discussed in this work.

Writers, reporters and citizens in China have argued that the Chinese textbooks currently used in mainland China are going in the wrong direction, and textbooks used $70$ years ago seem to be more reasonable. Influenced by English teaching, Chinese teaching indeed becomes increasingly speaking- and listening-oriented \cite{Lam2}. Speaking- and listening-oriented approach is a reasonable way to learn a phonetic language. However, for Chinese -- an ideographic language,  it results an inefficient learning order of Chinese character where structurally complicated characters are often taught before simpler ones. What we are suggesting is that in designing the speaking, listening and reading materials, one should utilize the logographic relations among Chinese characters and also respect  the optimal learning order discovered from analyzing the character network of the same relation. Only using a network analysis can we capture an entire picture of a network of these structural relations.

{\bf Acknowledgements} This work was supported by NSFC Grant $61174150$ and $60974084$.

{\bf Competing interests statement} The authors declare that they have no competing financial interests.

{\bf Correspondence} should be addressed to J. Wu (jinshanw@bnu.edu.cn).

\bibliographystyle{naturemag}
\bibliography{characters}

\section{Supporting Online Material}
\subsection{Data and methods}

\subsubsection{Decomposition of Chinese characters}
According to ``Liu Shu'' (six ways of creating Chinese characters), ideally when sub-characters are combined to form a character the compound character should be connected to its sub-characters either via their meanings or pronunciations. Thus, Chinese characters are usually meaningfully and coherently connected to each other. Let us start from the bottom of Fig. 1 in the main text. The four characters are ``\raisebox{-1mm}{\includegraphics[width=0.5cm]{char-one.pdf}}'' (one), ``\raisebox{-1mm}{\includegraphics[width=0.5cm]{char-big.pdf}}''(person, big), ``\raisebox{-1mm}{\includegraphics[width=0.5cm]{char-heart.pdf}}''(heart), ``\raisebox{-1mm}{\includegraphics[width=0.5cm]{char-water.pdf}}'' (water). These characters closely resemble the shapes or characteristics of the objects to which they refer, though their forms today might not hold as much of a resemblance as their ancient forms. One can compare the modern simplified Chinese character against their ancient Zhuanti forms in the figures. Such characters are called pictographic (Xiangxing) characters.

Initially, the character ``\raisebox{-1mm}{\includegraphics[width=0.5cm]{char-sky.pdf}}'' (sky) refers to the head, the primary part of a person, by placing a bar over the character ``\raisebox{-1mm}{\includegraphics[width=0.5cm]{char-big.pdf}}''(person, big). The meaning later developed and became the sky, heaven and god, \ie the primary part of everything as ancient Chinese people believed. This way of forming new characters from radical parts is called ``simple'' ideogram (Zhishi) or ``combination character'' ideogram (Huiyi). These two mechanisms are in fact slightly different in that the first is based on only one radical part, usually with only a very simple additional stroke while the second usually involves two radical parts. For a character formed by these two principles, its meaning usually can be read out intuitively from the combination. For example, the character `\raisebox{-1mm}{\includegraphics[width=0.5cm]{char-forest.pdf}}' (forest) mentioned in the introduction of the main text follows the principle of ``combined'' ideogram: it is a stack of three `\raisebox{-1mm}{\includegraphics[width=0.5cm]{char-tree.pdf}}'(tree). However, in this work, we will not distinguish the two mechanisms.

The character `\raisebox{-1mm}{\includegraphics[width=0.5cm]{char-shame.pdf}}' (, ashamed)  is a compound character of `\raisebox{-1mm}{\includegraphics[width=0.5cm]{char-sky.pdf}}' and `\raisebox{-1mm}{\includegraphics[width=0.5cm]{char-heart.pdf}}'. It follows a different principle, which later became popular in forming new Chinese characters, the so-called pictophonetic formation (Xingsheng). Here, `\raisebox{-1mm}{\includegraphics[width=0.5cm]{char-shame.pdf}}' and `\raisebox{-1mm}{\includegraphics[width=0.5cm]{char-sky.pdf}}' have exactly the same pronunciation, and the meaning of `\raisebox{-1mm}{\includegraphics[width=0.5cm]{char-shame.pdf}}' refers to a psychological phenomenon, which was believed to be related to `\raisebox{-1mm}{\includegraphics[width=0.5cm]{char-heart.pdf}}' (heart). The same pictophonetic relation holds among `\raisebox{-1mm}{\includegraphics[width=0.5cm]{char-add.pdf}}' (add), `\raisebox{-1mm}{\includegraphics[width=0.5cm]{char-shame.pdf}}'and `\raisebox{-1mm}{\includegraphics[width=0.5cm]{char-water.pdf}}' (water): the first two share the pronunciation while the last part `\raisebox{-1mm}{\includegraphics[width=0.5cm]{char-water.pdf}}' is remotely connected to the meaning of `\raisebox{-1mm}{\includegraphics[width=0.5cm]{char-add.pdf}}'.  In Fig.1 of the main text, we also notice that the characters `\raisebox{-1mm}{\includegraphics[width=0.5cm]{char-water.pdf}}' and `\raisebox{-1mm}{\includegraphics[width=0.5cm]{char-heart.pdf}}' also form the characters ` \raisebox{-2mm}{\includegraphics[width=0.5cm]{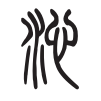}}'(seep). The character `\raisebox{-1mm}{\includegraphics[width=0.5cm]{char-qin.pdf}}' follows also the  pictophonetic formation. It is quite common that some basic characters are used in quite a few composed characters.

Here we have demonstrated four of the six principles. The other two are phonetic loan (Jiajie) and derivative cognates (Zhuanzhu). Those two principles are more on usage of characters but not on creating new characters. It is not our focus of this work to discuss various ways of usages of Chinese characters. Following the above general principles, our decompositions of characters  are based primarily on Ref.[11,12,21]. The first is a standard reference, where the six principles were first explicitly discussed, in Chinese etymology studies, and the last two are  regarded as developments of the first, mainly due to discoveries of new materials, including Oracle characters (Jiaguwen) and Bronze characters(Jinwen).

Starting from $3500$ characters, our network ends up with a giant component of $3687$ nodes and $7025$ links, plus $15$ isolated nodes. Why do we have more nodes than the total number of characters we start with? In our decomposition, we find some sub-characters beyond the set of the most used $3500$ characters. Sometimes, such sub-characters are just variations of their normal forms. The situation becomes more complicated when a radical whose corresponding normal form is not within the most-used set. In such cases, we add the ``never-independent characters'' as extra nodes in the network. For example, `\raisebox{-1mm}{\includegraphics[width=0.5cm]{char-shame.pdf}}' is such a rarely used character, but we keep it in our network.

See Fig. 2 in the main text for the full map of structural relations among Chinese characters.

\subsubsection{Additional explanation of definition of learning cost}
We define the learning cost of a character for a student to be the sum of the number of sub-characters and the learning cost (calculated recursively) of the unlearned sub-characters at his current stage. The recursive definition seems to imply that when a student is learning a compound character,  he has to recognize first the sub-characters. However, the dynamic process is only a fictitious process used to represent the difficulty that the student faces in learning the character. It does not means the learning process is indeed as such. Recall from the main text total cost of learning `\raisebox{-1mm}{\includegraphics[width=0.5cm]{char-add.pdf}}' before `\raisebox{-1mm}{\includegraphics[width=0.5cm]{char-shame.pdf}}' is $4=2+2$, which is from the fact that it has $2$ sub-characters and also from the fact that cost of learning the unknown `\raisebox{-1mm}{\includegraphics[width=0.5cm]{char-shame.pdf}}' is $2$. Therefore, determining cost of learning `\raisebox{-1mm}{\includegraphics[width=0.5cm]{char-add.pdf}}' first obviously involves cost of learning `\raisebox{-1mm}{\includegraphics[width=0.5cm]{char-shame.pdf}}'. However, this does not imply that the student should have known `\raisebox{-1mm}{\includegraphics[width=0.5cm]{char-shame.pdf}}' after acquiring `\raisebox{-1mm}{\includegraphics[width=0.5cm]{char-add.pdf}}'. If it happens so that the next time the student must learn `\raisebox{-1mm}{\includegraphics[width=0.5cm]{char-shame.pdf}}', then the learning cost of `\raisebox{-1mm}{\includegraphics[width=0.5cm]{char-shame.pdf}}' is still $2$ even he had learned `\raisebox{-1mm}{\includegraphics[width=0.5cm]{char-add.pdf}}' before. Thus the total learning cost of the two characters following the order of  `\raisebox{-1mm}{\includegraphics[width=0.5cm]{char-add.pdf}}' $\rightarrow$ `\raisebox{-1mm}{\includegraphics[width=0.5cm]{char-shame.pdf}}' is $6$.

Of course, if the student learned the character  `\raisebox{-1mm}{\includegraphics[width=0.5cm]{char-add.pdf}}' meaningfully, \ie when he learn the character  ` \raisebox{-1mm}{\includegraphics[width=0.5cm]{char-add.pdf}}', he indeed learn also the relation between  `\raisebox{-1mm}{\includegraphics[width=0.5cm]{char-add.pdf}}' and ` \raisebox{-1mm}{\includegraphics[width=0.5cm]{char-shame.pdf}}' (also the meaning of `\raisebox{-1mm}{\includegraphics[width=0.5cm]{char-shame.pdf}}') explicitly from his books or his instructors,  then the total cost for him to learn both characters is in fact $4$ (no cost for learning `\raisebox{-1mm}{\includegraphics[width=0.5cm]{char-shame.pdf}}'), which is the same cost of learning both characters in the order of `\raisebox{-1mm}{\includegraphics[width=0.5cm]{char-shame.pdf}}' $\rightarrow$ ` \raisebox{-1mm}{\includegraphics[width=0.5cm]{char-add.pdf}}'. Therefore, learning closely connected characters together at the same time and learning them meaningfully would reduce the cost. Therefore, one might conclude that our definition of learning cost does not apply to such meaningful learning. However, for this we would argue that such meaningful learning has implicitly used the optimal learning orders, learning the two characters simultaneously and meaningfully is equivalent to learning them according to the proper order.

Another problem related to our definition of learning cost is that we treated the number of sub-characters and the cost of unlearned sub-characters equally. This can be questioned and should be investigated further. For example, one might introduce a parameter to rescale the number of sub-characters  and then sum the two together. For simplicity, we have not yet discussed this issue. Finding the proper value of such parameters from empirical studies and then comparing performance of those learning orders again using the new definition of cost should be an interesting topic.

\subsection{Supplemental Results}
At last, we provide the two important lists of characters as final results of our network-based analysis of Chinese characters. First is the adjacency list of the network of characters. The first character of every line is the starting point of links and all other characters in the same line are the ending point of the links, meaning the first character is a part of everyone of the other characters. Second is the order of Chinese characters listed according to the calculated DNW centrality. This list includes all $3500$ characters and $b=0.5$ is used in the calculation of DNW. In the main text, when $1775$ characters are used as the learning target, we find the optimal value of parameter $b$ is $b=0.35$. Repeating the same analysis for all $3500$ characters, we find that learning efficiency is higher when $b=0.5$ is used instead of $b=0.35$.  Here the list is produced when we consider the whole set of most used characters as the learning goal. The lists can be downloaded from our own still developing website on Chinese learning \href{http://www.learnm.org/data/}{http://www.learnm.org/data/}.

\end{document}